\documentstyle[12pt]{article}
\newcommand{\svec}[1]{ \stackrel{\rightarrow}{#1} }

\newcommand{\ehat}{ \hat U_{\epsilon} }

\newcommand{\define}{ \stackrel{\triangle}{=} }

\def\be{\begin{equation}}
\def\ee{\end{equation}}
\def\ba{\begin{array}}
\def\ea{\end{array}}
\begin{document}
\title{\bf Quantum Gauge Theory of Gravity }
\author{{Ning Wu}
\thanks{email address: wuning@ihep.ac.cn}  \\
{\small Institute of High Energy Physics, P.O.Box 918-1,
Beijing 100039, P.R.China} }
\maketitle
\vskip 0.8in

~~

\vskip 0.8in

\begin{abstract}
The quantum gravity is formulated based on principle of
local gauge invariance.  The model discussed in this paper
has local gravitational gauge symmetry and gravitational field
is represented by gauge field. In leading order approximation,
it gives out classical Newton's theory of gravity. It can also
give  out an Einstein-like field equation with cosmological 
constant. For classical tests, it gives out the same theoretical
predictions as those of general relativity. This quantum gauge
theory of gravity is a renormalizable quantum theory.
\\

\end{abstract}

\newpage
\Roman{section}

\section{Introduction}
\setcounter{equation}{0}

Gravity is an ancient topic in science. In ancient times, human has
known the existence of weight, which is the  gravity between an
object and earth. In 1686, Isaac Newton
published his epoch-making book {\it MATHEMATICAL
PRINCIPLES OF NATURAL PHILOSOPHY}.
Through studying the motion of planet in solar system,
he found that the gravity obeys the inverse square law\cite{01}.
The Newton's classical theory of gravity is kept unchanged
until 1916. At that year, Einstein published his
epoch-making paper on General Relativity\cite{02,03}. In this
great work, he founded a relativistic theory on gravity, which
is based on principle of general relativity and equivalence
principle. Newton's classical theory for gravity appears as
a classical limit of general relativity.\\

1954, Yang and Mills proposed non-Abel gauge field
theory\cite{1}. This theory was soon applied to elementary
particle physics. Unified electroweak theory\cite{2,3,4}
and quantum chromodynamics are all based on gauge field theory.
Now it is generally  believed that four kinds of
fundamental interactions in Nature are all gauge
interactions and they can be described by gauge field theory.
From theoretical point of view, the principle of local
gauge invariance plays a fundamental role in particle's
interaction theory.  \\

In 1916, Albert Einstein points out that quantum effects must
lead to modifications in the theory of general relativity\cite{5}.
Soon after the foundation of quantum mechanics,
physicists try to found a theory that could describe the
quantum behavior of the full gravitational field. In the 70
years attempts, physicists have founded two theories based
on quantum mechanics that attempt to unify general
relativity and quantum mechanics, one is canonical quantum
gravity and another is superstring theory. But for quantum
field theory, there are different kinds of mathematical
infinities that naturally occur in quantum descriptions of
fields. These infinities should be removed by the technique
of perturbative renormalization. However, the perturbative
renormalization does not work for the quantization of Einstein's
theory of gravity, especially in canonical quantum gravity. In
superstring theory, in order to make perturbative renormalization
to work, physicists have to introduce six extra dimensions. But
up to now, none of the extra dimensions have been observed.
To found a consistent theory that can unify general relativity and
quantum mechanics is a long dream for physicists. \\

Gauge treatment of gravity was suggested immediately after
the gauge theory birth itself\cite{b1,b2,b3,b31,b32,b33}. 
In the traditional
gauge treatment of gravity, Lorentz group is localized, and the
gravitational fields are not represented by gauge
potentials\cite{b4,b5,b6}. The theory has beautiful mathematical
forms, but it is considered to be non-renormalizable. \\

\section{Gauge Principle}
\setcounter{equation}{0}

In this paper, we will use completely new notion and completely
new method to study quantum gravity. Our goal is to set up a
consistent quantum gauge theory of gravity which is
renormalizable. The foundation of the new quantum gauge theory
of gravity is gauge principle.  Gauge principle can
be formulated as follows: Any kind of fundamental interactions
has a gauge symmetry corresponding to it; the gauge
symmetry completely determines the forms of interactions.
In principle, gauge principle has the following
four different contents:
\begin{enumerate}

\item {\bf Conservation Law:} the global gauge symmetry
gives out conserved current and conserved charge;

\item {\bf Interactions:} the requirement of the local gauge
symmetry requires introduction of gauge field or a set of gauge
fields; the interactions between gauge fields and matter fields are
completely determined by the requirement of local gauge symmetry;
these gauge fields transmit the corresponding interactions;

\item {\bf Source:} qualitative speaking, the conserved charge
given by global gauge symmetry is the source of gauge field;
for non-Abel gauge field, gauge field is also the source of itself;

\item {\bf Quantum Transformation:} the conserved charges given
by global gauge symmetry become generators of quantum
gauge transformations after quantization, and for this kind of
of interactions, the quantum transformations can not have
generators other than quantum conserved charges given by
global gauge symmetry.

\end{enumerate}
Gauge principle tells us how to study conservation law and
fundamental interactions through symmetry.  \\

It is known that the source of gravitational interactions is
energy-momentum, and energy-momentum is the conserved charge of
global space-time translation. According to gauge principle,
space-time translation group is the symmetry group corresponding
to gravity. But the traditional space-time translation group
consists of those transformations that objects and fields are kept
unchanged while space-time coordinates undergo some
transformations. In gravitational gauge theory, we need another
kind of translation, that is, space-time coordinates are kept
unchanged while objects and fields undergo some transformations.
This kind of transformation is called gravitational gauge
transformation. From mathematical point of view, gravitational
gauge transformation is the inverse transformation of space-time
translation, so they are essentially the same. But from physical
point of view, they are not the same, especially when we discuss
gravitational gauge field. In a meaning, space-time translation is
a kind of mathematical transformation; while gravitational gauge
transformation is a kind of physical transformation, which
contains dynamical information of interactions.
\\

In traditional gauge theory of gravity, Lorentz group is localized.
The generators of Lorentz group are angular momentum tensor
$M_{\mu \nu}$. According to gauge principle, if we localize
Lorentz group, the theory will contain spin-spin interactions,
which do not belong to traditional Newton-Einstein gravity.
It is known that a theory which contain direct spin-spin
interactions is not renormalizable. On the other hand, we do not
find any direct evidence on the existence of the direct spin-spin
interactions in Nature. For these reasons, we will not localize
Lorentz group in this paper. We only localize gravitational gauge
group, which is enough for us to set up a consistent quantum
theory of gravity. Besides, the gravitational field is represented
by gauge potential, and space-time is kept flat.
But, if we go into the gravitational gauge group space,
we will find that this space-time is curved. However, foundations
of gauge theory of gravity is not formulated in this space.
\\

\section{Gravitational Interactions of Pure Gravitational Gauge
Field}
\setcounter{equation}{0}

Gravitational gauge group is a transformation group which
consists of all non-singular translation operators $\ehat$, where
$\ehat$ is given by
\be  \label{3.1}
\ehat \define  E^{- i \epsilon^{\mu} \cdot \hat{P}_{\mu}},
~~~~~
\hat{P}_{\mu} = -i \frac{\partial}{\partial x^{\mu}}.
\ee
In eq.(\ref{3.1}), $E^{a^{\mu} \cdot b_{\mu}}$ is a special
exponential function whose definition is
\be  \label{3.2}
E^{a^{\mu} \cdot b_{\mu}} \define 1 +
\sum_{n=1}^{\infty} \frac{1}{n!}
a^{\mu_1} \cdots a^{\mu_n} \cdot
b_{\mu_1} \cdots b_{\mu_n}.
\ee
The gravitational gauge transformation of scalar field and
vector field respectively are,
\be  \label{3.3}
\phi(x) \to \phi'(x) = ( \ehat \phi(x)).
\ee
\be  \label{3.4}
A_{\mu}(x) \to A'_{\mu}(x) = ( \ehat A_{\mu}(x)).
\ee
A vector field can be a scalar or a vector or a tensor in
the space of gauge group.
If $A_{\mu}(x)$ is a vector in the space of
gravitational gauge group, it can be expanded as:
\be  \label{3.5}
A_{\mu}(x)  =  A_{\mu}^{\alpha}(x) \cdot \hat{P}_{\alpha},
\ee
where index $\alpha$ is group index and index $\mu$ is ordinary
Lorentz index. The transformation of component field is
\be  \label{3.6}
A_{\mu}^{\alpha}(x)  \to A_{ \mu}^{\prime\alpha}(x) =
\Lambda^{\alpha}_{~~\beta} \ehat A_{\mu}^{\beta}(x) \ehat^{-1} ,
\ee
where $\Lambda^{\alpha}_{~~\beta} $ is given by
\be  \label{3.7}
\Lambda^{\alpha}_{~~\beta} =
\frac{\partial x^{\alpha}}{\partial ( x - \epsilon (x) )^{\beta}},
\ee
$A_{\mu}(x)$ can also be a $n$th order tensor in the space of
gravitational gauge group.
Under gravitational gauge transformation, the behavior
of a group index is quite different from that of a 
Lorentz index. However, they have the same behavior 
under global Lorentz transformation. \\

Define matrix $G$ as 
\be  \label{3.701} 
G = (G_{\mu}^{\alpha}) = (
\delta_{\mu}^{\alpha} - g C_{\mu}^{\alpha}), 
\ee 
where
$C_{\mu}^{\alpha}$ is the gravitational gauge field 
which will be introduced below.  A simple form for 
matrix $G$ is 
\be
\label{3.702} 
G = I - gC, 
\ee 
where $I$ is a unit matrix and $C= (C_{\mu}^{\alpha})$. 
Therefore, 
\be  \label{3.703} 
G^{-1} =
\frac{1}{I-gC}. 
\ee 
$G^{-1}$ is the inverse matrix of $G$, it
satisfies 
\be  \label{3.704} 
(G^{-1})^{\mu}_{\beta} G^{\alpha}_{\mu} 
= \delta_{\beta}^{\alpha}, 
\ee 
\be  \label{3.705}
G^{\alpha}_{\mu} (G^{-1})^{\nu}_{\alpha} 
= \delta^{\nu}_{\mu}. 
\ee
Define 
\be  \label{3.706} 
g^{\alpha \beta} \define \eta^{\mu \nu}
G_{\mu}^{\alpha} G_{\nu}^{\beta}, 
\ee 
\be  \label{3.707} 
g_{\alpha \beta} \define \eta_{\mu \nu} 
(G^{-1})_{\alpha}^{\mu} (G^{-1})_{\beta}^{\nu}. 
\ee 
It can be easily proved that 
\be \label{3.708} 
g_{\alpha \beta} g^{\beta \gamma} =
\delta_{\alpha}^{\gamma}, 
\ee 
\be  \label{3.708} 
g^{\alpha \beta} g_{\beta \gamma} 
= \delta^{\alpha}_{\gamma}. 
\ee 
Under gravitational gauge transformations, they transform as 
\be
\label{3.709} 
g_{\alpha \beta}(x)  \to g'_{\alpha \beta}(x') =
\Lambda_{\alpha}^{~\alpha_1} 
\Lambda_{\beta}^{~\beta_1} 
(\ehat g_{\alpha_1 \beta_1}(x)), 
\ee 
\be  \label{3.710} 
g^{\alpha \beta}(x)  
\to g^{\prime \alpha \beta}(x') =
\Lambda^{\alpha}_{~\alpha_1} 
\Lambda^{\beta}_{~\beta_1} (\ehat
g^{\alpha_1 \beta_1}(x)), 
\ee
\\

The gravitational gauge covariant derivative is defined by
\be  \label{3.10}
D_{\mu} = \partial_{\mu} - i g C_{\mu} (x),
\ee
where $C_{\mu} (x)$ is the gravitational gauge field and $g$ is the
gravitational gauge coupling constant. It is a Lorentz
vector. Under gravitational gauge transformation, $C_{\mu} (x)$
 transforms as
\be  \label{3.11}
C_{\mu}(x) \to  C'_{\mu}(x) =
\ehat (x) C_{\mu} (x) \ehat^{-1} (x)
+ \frac{i}{g} \ehat (x) (\partial_{\mu} \ehat^{-1} (x)),
\ee
and $D_{\mu}$ transforms covariantly,
\be  \label{3.12}
D_{\mu} (x) \to D'_{\mu} (x)
= \ehat D_{\mu} (x) \ehat^{-1}.
\ee
Gravitational gauge field $C_{\mu} (x)$ is a vector field, it is a
Lorentz vector. It is also a vector in gauge group space, so it can be
expanded as linear combinations of generators of gravitational gauge
group:
\be  \label{3.13}
C_{\mu} (x) = C_{\mu}^{\alpha}(x) \cdot \hat{P}_{\alpha}.
\ee
$C_{\mu}^{\alpha}$ is component field of gravitational gauge
field. It looks like a second rank tensor, but it is not tensor
field. The index $\alpha$ is not an ordinary Lorentz index,
it is just a gauge
group index. Gravitational gauge field $C_{\mu}^{\alpha}$ has only
one Lorentz index, so it is a kind of vector field. \\

The strength of gravitational gauge field is defined by
\be  \label{3.14}
F_{\mu \nu} = \frac{1}{-i g}
\lbrack D_{\mu} ~~,~~ D_{\nu} \rbrack,
\ee
or
\be  \label{3.15}
F_{\mu \nu} = \partial_{\mu} C_{\nu}(x)
- \partial_{\nu} C_{\mu}(x)
- i g C_{\mu}(x) C_{\nu}(x)
+ i g  C_{\nu}(x) C_{\mu}(x).
\ee
$F_{\mu \nu}$ is a second rank Lorentz tensor. It is a vector is
group space, so it can be expanded in group space,
\be  \label{3.16}
F_{\mu \nu} (x) = F_{\mu \nu}^{\alpha} (x) \cdot \hat{P}_{\alpha}.
\ee
The explicit form of component field strength is
\be  \label{3.17}
F_{\mu \nu}^{\alpha} = \partial_{\mu} C_{\nu}^{\alpha}
- \partial_{\nu} C_{\mu}^{\alpha}
- g C_{\mu}^{\beta} \partial_{\beta} C_{\nu}^{\alpha}
+ g C_{\nu}^{\beta} \partial_{\beta} C_{\mu}^{\alpha}
\ee
Under gravitational gauge transformation, The component
strength of gravitational gauge field transforms as
\be  \label{3.1701}
F_{\mu \nu}^{\alpha} \to F_{\mu \nu}^{\prime\alpha}
= \Lambda^{\alpha}_{~\beta} (\ehat F_{\mu \nu}^{\beta}).
\ee
\\

Similar to traditional gauge field theory, the kinematical 
term for gravitational gauge field can be selected as
\be  \label{3.18}
{\cal L}_0 = - \frac{1}{4} \eta^{\mu \rho} \eta^{\nu \sigma}
g_{\alpha \beta}
F_{\mu \nu}^{\alpha} F_{\rho \sigma}^{\beta},
\ee
where $\eta^{\mu \rho}$ is the Minkowski metric.
Using eq.(\ref{3.709}) and eq.(\ref{3.1701}).
We can easily prove that this Lagrangian does 
not invariant under gravitational gauge transformation, 
it transforms covariantly
\be  \label{3.19}
{\cal L}_0 \to {\cal L}'_0 = ( \ehat {\cal L}_0).
\ee
In order to resume the gravitational gauge symmetry of 
the action, we introduce an  important factor, which is 
denoted as $J(C)$. The form of $J(C)$ is not 
unique\cite{wu1}. In this paper, $J(C)$ is selected to be
\be  \label{3.201}
J(C) = \sqrt{ - {\rm det} (g_{\alpha\beta})},
\ee
where $g_{\alpha\beta}$ is given by eq.(\ref{3.707}). Using
eq.(\ref{3.709}), we can
prove that under gravitational gauge transformation, $J(C)$
transforms as
\be  \label{3.202}
J(C) \to J'(C') = J \cdot (\ehat J(C)),
\ee
where
\be  \label{3.203}
J \define {\rm det} (\Lambda_{\alpha}^{~\beta})
= {\rm det} \left(\frac{\partial(x - \epsilon)^{\mu}}
{\partial x^{\nu}} \right),
\ee
which is the Jacobian of the corresponding transformation\cite{wu1}.
The Lagrangian for gravitational gauge field is selected as
\be  \label{3.21}
{\cal L} \define J(C) \cdot {\cal L}_0,
\ee
and the action for gravitational gauge field is
\be  \label{3.22}
S = \int {\rm d}^4 x {\cal L}.
\ee
Using the following identity
\be  \label{3.2201}
\int {\rm d}^4 x J \cdot (\ehat f(x))
= \int {\rm d}^4 x f(x),
\ee
and eq.(\ref{3.19}) and eq.(\ref{3.202}), we can
prove that this action has local gravitational gauge symmetry.
\\

According to gauge principle, the global symmetry gives out
conserved current, it is
\be  \label{3.23}
T_{i \alpha}^{\mu} \define  J(C)
\left(- \frac{\partial {\cal L}_0}
{\partial \partial_{\mu} C_{\nu}^{\beta}}
\partial_{\alpha} C_{\nu}^{\beta}
+ \delta^{\mu}_{\alpha} {\cal L}_0 \right).
\ee
We call it inertial energy-momentum tensor.
The equation of motion of gravitational gauge field is
\be  \label{3.24}
\partial_{\mu} (\eta^{\mu \lambda} \eta^{\nu \tau} g_{\alpha \beta}
F_{\lambda \tau}^{\beta} )  = - g T_{g \alpha}^{\nu}.
\ee
$T_{g \alpha}^{\nu}$ is also a conserved current. We call it
gravitational energy-momentum tensor, which is the source of
gravitational gauge field. Its explicit form is
\be  \label{3.25}
\ba{rcl}
T_{g \alpha}^{\nu} & = &
 - \frac{\partial {\cal L}_0}{\partial D_{\nu} C_{\mu}^{\beta}}
\partial_{\alpha} C_{\mu}^{\beta}
+ G_{\alpha}^{-1\nu} {\cal L}_0
-  G^{-1 \lambda}_{\sigma}
(\partial_{\mu} C_{\lambda}^{\sigma})
\frac{\partial {\cal L}_0}
{\partial \partial_{\mu} C_{\nu}^{\alpha}} \\
&& \\
&&- \frac{1}{2} \eta^{\mu\rho} \eta^{\lambda\sigma}
g_{\alpha\beta} G^{-1\nu}_{\gamma}
F^{\gamma}_{\mu\lambda} F^{\beta}_{\rho\sigma}
+ \partial_{\mu} (\eta^{\nu \lambda}
\eta^{\sigma \tau}  g_{ \alpha\beta}
F_{\lambda \tau}^{\beta} C_{\sigma}^{\mu}).
\ea
\ee
Now, we have obtained two different energy-momentum tensors, one
is inertial energy-momentum tensor, another is gravitational
energy-momentum tensor. The conserved charge given by inertial
energy-momentum tensor is the inertial energy-momentum, and the
conserved charge given by gravitational energy-momentum tensor is the
gravitational energy-momentum.
\\

This quantum gauge theory of gravity is a renormalizable quantum
theory. A detailed and strict proof on the renormalizability of the
theory can be found in ref. \cite{wu1}. We will not discuss this
problem here.\\

\section{Classical Limit of the Theory}
\setcounter{equation}{0}

Now, let's see the classical limit of the above theory. Suppose
that the gravitational field is static and weak, and the moving speeds
of all objects are slow. Then, in leading order approximation,
eq.(\ref{3.24}) gives out
\be  \label{4.1}
\nabla ^2 C_0^0 = - g T^0_0.
\ee
Suppose that the source of gravitational field is a point object,
that is
\be  \label{4.2}
T^0_0 = - M \delta(\svec{x}) .
\ee
From eq.(\ref{4.1}), we can obtain that
\be  \label{4.3}
gC^0_0  = - \frac{g^2 M}{4 \pi r}.
\ee
This is just the gravitational potential which is expected in Newton's
theory of gravity. Therefore, in leading order approximation, this theory
gives out classical Newton's theory of gravity.\\

\section{Einstein-like Field Equation With Cosmological Constant}
\setcounter{equation}{0}

In the above chapters, the quantum gravity is formulated
in the traditional framework of quantum field theory, i.e., 
the physical space-time is always flat and the space-time
metric is always selected to be the Minkowski metric. In this 
picture, gravity is treated as physical interactions 
in flat physical space-time. Our gravitational gauge
transformation does not act on physical space-time
coordinates, but act on
physical fields, so gravitational gauge transformation
does not affect the structure of physical space-time.
This is one picture of gravity, or call it  one 
representation of gravity theory. For the sake of simplicity,
we call it physical representation of gravity.\\

There is another representation of gravity which is widely
used in Einstein's general relativity. This representation
is essentially a geometrical representation of gravity. 
For gravitational gauge theory, if we treat 
$G^{\alpha}_{\mu}$ ( or $G^{-1 \mu}_{\alpha}$ )
as a fundamental physical input of the
theory, we can also set up the geometrical representation
of gravity\cite{wu21}. For gravitational gauge theory, the 
geometrical nature of gravity essentially originates from the 
geometrical nature of the gravitational gauge transformation.
In the geometrical picture of gravity, gravity is not 
treated as a kind of physical interactions, but is
treated as the geometry of space-time. So, there 
is no physical gravitational interactions in space-time
and space-time is curved if 
gravitational field does not vanish. In this case, the
space-time metric is not Minkowski metric, but 
$g^{\alpha\beta}$ (or $g_{\alpha\beta}$). In other words,
Minkowski metric is the space-time metric if we discuss
gravity in physical representation while metric tensor
$g^{\alpha\beta}$ ( or $g_{\alpha\beta}$) is space-time
metric if we discuss gravity in geometrical representation.
So, if we use Minkowski metric in our discussion, 
it means that we are in physical representation of 
gravity; if we use metric tensor 
$g^{\alpha\beta}$ (or $g_{\alpha\beta}$) in our discussion,
it means that we are in geometrical representation. \\

In one representation, gravity is treated as physical
interactions, while in another representation, gravity 
is treated as geometry of space-time. But we know that
there is only one physics for gravity, so two 
representations of gravity must be equivalent. This
equivalence means that the nature of gravity is
physics-geometry duality, i.e., gravity is a kind 
of physical interactions which has the characteristics 
of geometry; it is also a geometry of space-time 
which is essentially a kind of physical interactions.
Now, let's go into the geometrical representation of
gravity and use $g^{\alpha\beta}$ and $g_{\alpha\beta}$ 
as space-time metric tensors. In this picture, we can
obtain an Einstein-like field equation with 
cosmological constant. \\ 

Define
\be  \label{50.1}
\Lambda \define \frac{1}{2} ( R + 4 g^2 {\cal L}_0).
\ee
Then action given by eq.(\ref{3.22}) will be changed into
\be \label{50.2}
S = S_g + S_M,
\ee
where
\be  \label{50.3}
S_g = \frac{-1 }{16 \pi G} \int {\rm d}^4x
\sqrt{- g} (R - 2 \Lambda),
\ee
\be  \label{50.4}
S_M = \int {\rm d}^4x {\cal L}_M,
\ee
where $G$ is the Newtonian gravitational constant, which
is given by
\be  \label{50.44}
G = \frac{g^2}{4 \pi},
\ee
$R$ is the scalar curvature, $\Lambda$ is the cosmological
constant, ${\cal L}_M$ is the lagrangian density for
matter fields. Scalar curvature $R$ can be expressed
by gravitational gauge field $C_{\mu}^{\alpha}$\cite{wu1}.
We have added the action for matter
fields into eq.(\ref{50.2}) and denoted the action for
pure gravitational gauge field as $S_g$.
 Using the following relations
\be \label{50.5}
\delta \sqrt{-g} = \frac{1}{2}
\sqrt{-g } g^{\mu\nu} \delta g_{\mu\nu},
\ee
\be \label{50.6}
\sqrt{-g } g^{\mu\nu} \delta R_{\mu\nu}
= \partial_{\lambda} W^{\lambda},
\ee
\be \label{50.7}
T_m^{\mu\nu} = \frac{2}{\sqrt{-g }}
\frac{\delta S_M}{\delta g_{\mu\nu}(x)},
\ee
where $T_m^{\mu\nu}$ is the energy-momentum tensor of
matter fields and $W^{\lambda}$ is a  contravariant
vector, we can obtain the Einstein's field equation
with cosmological constant $\Lambda$,
\be \label{50.8}
R_{\mu\nu} - \frac{1}{2} g_{\mu\nu} R
+ \Lambda g_{\mu\nu}
= - 8 \pi G T_{\mu\nu},
\ee
where $T_{\mu\nu}$ is the revised energy-momentum tensor,
whose definition is
\be \label{50.9}
T_{\mu\nu} \define T_{m \mu\nu}
- \frac{1}{4 \pi G}
\frac{\delta \Lambda}{\delta g^{\mu\nu}}.
\ee
In eq.(\ref{50.9}), the difinition of 
$\frac{\delta \Lambda}{\delta g^{\mu\nu}}$
is not clear, because $\Lambda$ is a function of 
$G^{\alpha}_{\mu}$, not a function of 
$g^{\mu \nu}$. So, we need to give out an explicite
definition of $\frac{\delta \Lambda}{\delta g^{\mu\nu}}$.
According to eq.(\ref{3.706}), we have
\be \label{50.10}
\frac{\partial g^{\mu\nu}}
{\partial G^{\lambda}_{\alpha}}
= \delta^{\mu}_{\lambda} 
\eta^{\alpha\beta} G^{\nu}_{\beta}
+  \delta^{\nu}_{\lambda} 
\eta^{\alpha\beta} G^{\mu}_{\beta}.
\ee 
Therefore, we have
\be \label{50.11}
\frac{\delta \Lambda}{\delta G^{\lambda}_{\alpha}}
= 2 \eta^{\alpha \beta} G^{\nu}_{\beta}
\frac{\delta \Lambda}{\delta g^{\lambda\nu}}
\ee
It gives out
\be \label{50.12}
\frac{\delta \Lambda}{\delta g^{\mu\nu}} 
= \frac{1}{4} \eta_{\alpha \beta} 
\left( G_{\nu}^{-1 \beta}
\frac{\delta \Lambda}{\delta G^{\mu}_{\alpha}}
+  G_{\mu}^{-1 \beta}
\frac{\delta \Lambda}{\delta G^{\nu}_{\alpha}}
\right), 
\ee
which gives out the explicite expression for 
$\frac{\delta \Lambda}{\delta g^{\mu\nu}}$.
Eq.(\ref{50.8}) is quite like the Einstein's field
equation with cosmological constant in form, so
we call it the Einstein-like field equation
with cosmological constant.
\\

The explicite expression of ${\cal L}_0$ is given
by eq.(\ref{3.18}), and the explicite expression of
scalar curvature $R$ can be found in \cite{wu1}.
According to eq.(\ref{50.1}), the explicate formulation
of $\Lambda$ can be calculated. In other words, the
cosmological constant can be expressed in terms of
gravitational gauge field $C_{\mu}^{\alpha}$.
The we make large scale average of it which will gives
out the average value of the cosmological constant.
A rough estimation shows that\cite{wu3}
\be  \label{50.10}
\Lambda \sim 2.92 \times 10^{-52} m^{-2},
\ee
which is quite close to experimental results\cite{pdg},
\be  \label{50.11}
\Lambda = 3.51 \times 10^{-52} \Omega_{\Lambda} h_0^2 m^{-2},
\ee
with $\Omega_{\Lambda}$ is the scaled cosmological constant
and $h_0$ is the normalized Hubble expansion rate, whose
values are
\be  \label{50.12}
-1< \Omega_{\Lambda} <2,
\ee
\be  \label{50.13}
0.6 < h_0 < 0.8.
\ee

\section{Classical Tests}
\setcounter{equation}{0}

It is known that General Relativity is tested by three
main classical tests, which are  perihelion procession,
deflection of light and gravitational red-shift. All these
three tests are related to geodesic curve equation and
schwarzachild solution in general relativity. 
If we know geodesic curve equation and space-time
metric, we can calculate perihelion procession,
deflection of light and gravitational red-shift. 
In this chapter, we discuss this problem
from the point of view of gauge theory of gravity.
\\

In order to discuss classical tests of gravity, for 
the sake of convience, we use the geometrical 
representation of gravity. As we have stated above, 
in the geometrical representation of gravity, 
gravity is not treated as physical interactions 
in space-time. In the gemetrical representation
of gravity, we use the same manner which is widely
used in general relativity to discuss classical
tests and to explain the predictions with
observations. In the geometrical representation of
gravity, if there is no other
physical interactions in space time, any mass point
can not feel any physical forces when it moves in 
space-time. So, it must move along the curve which
has the least length. 
Suppose that a particle is moving from point
A to point B along a curve. Define
\be  \label{6.01}
T_{BA} = \int_A^B 
\sqrt{ - g_{\mu\nu} \frac{{\rm d}x^{\mu}}{{\rm d}p}
\frac{{\rm d}x^{\nu}}{{\rm d}p} } {\rm d} p,
\ee
where $p$ is a parameter that describe the orbit.
The real curve that the particle moving along 
corresponds to the minimum of $T_{AB}$. The
minimum of $T_{AB}$ gives out the following
geodesic curve equation
\be  \label{6.02}
\frac{{\rm d}^2 x^{\mu}}{{\rm d}p^2}
+ \Gamma^{\mu}_{\nu\lambda}
\frac{{\rm d}x^{\nu}}{{\rm d}p} 
\frac{{\rm d}x^{\lambda}}{{\rm d}p} =0,
\ee
where $\Gamma^{\mu}_{\nu\lambda}$ is the affine 
connection
\be
\Gamma^{\lambda}_{\mu \nu}
= \frac{1}{2} g^{\lambda \sigma}
\left( \frac{\partial g_{\mu \sigma}}{\partial x^{\nu}}
+ \frac{\partial g_{\nu \sigma}}{\partial x^{\mu}}
-\frac{\partial g_{\mu \nu}}{\partial x^{\sigma}} \right).
\label{6.03}
\ee
Eq.(\ref{6.02}) gives out the curve that a free particle 
moves along in curved space-time if we discuss physics
in the geometrical representation of gravity.
\\

Now, we need to calculate a schwarzchild-like solution
in gauge theory of gravity. In chapter 4,
we have obtained a solution of $C_{\mu}^{\alpha}$
for static spherically symmetric gravitational fields
in linear approximation of $gC_{\mu}^{\alpha}$. But
expetimental tests, especially  perihelion procession,
are sensitive to second order of $gC_{\mu}^{\alpha}$.
The best way to do this is to solve the equation of
motion of gravitational gauge field in the second
order approximation of $gC_{\mu}^{\alpha}$. But this
equation of motion is a non-linear second order
partial differential equations. It is rather difficult
to solve them. So, we had to find some other method
to do this.  The perturbation method is used to do
this.  After considering
corrections from gravitational energy of the sun in vacuum
space and gravitational interactive energy between the sun and
the Mercury, the equivalent gravitational gauge field
in the second order approximation is\cite{wu4}
\be  \label{6.1}
g C_{0}^{0} = - \frac{GM}{r}
- \frac{3}{2} \frac{G^2 M^2}{r^2}
+ O\left(\frac{G^3 M^3}{r^3} \right).
\ee
Then using eq.(\ref{3.707}), we can obtain the following
solution
\be  \label{6.2}
\ba{rcl}
d \tau^2 &=& \left\lbrack
1 - \frac{2 GM}{r} + O\left(\frac{G^3 M^3}{r^3} \right)
\right\rbrack d t^2
-  \left\lbrack
1 + \frac{2 GM}{r} + O\left(\frac{G^2 M^2}{r^2} \right)
\right\rbrack d r^2 \\
&&\\
&& - r^2 d \theta^2 - r^2 {\rm sin}^2\theta d \varphi^2,
\ea
\ee
where we have use the following gauge for gravitational
gauge field,
\be  \label{6.3}
C_{\mu}^{\mu} = 0.
\ee
This solution is quite similar to schwarzschild solution,
but it is not schwarzschild solution, so we call it
schwarzschild-like solution.
If we use Eddington-Robertson expansion, we will find that
for the present schwarzschild-like solution\cite{wei},
\be  \label{6.4}
\alpha = \beta = \gamma = 1.
\ee
They have the same values as schwarzschild solution in
general relativity and all three tests are only sensitive
to these three parameters,  so gauge theory of gravity
gives out the same  theoretical predictions as those of
general relativity\cite{wu4}. More
detailed discussions on classical tests can be found
in literature \cite{wu4}. (This result hold for those
models which have other choice
of $\eta_2$ and $J(C)$ which is duscussed in \cite{wu1}.)\\

\section{Summary and Comments}
\setcounter{equation}{0}

In this paper, we have discussed a completely new quantum gauge
theory of gravity. Finally, we give a simple summary to the
whole theory.
1) In leading order approximation, the gravitational gauge field
theory gives out classical Newton's theory of gravity.
2) It gives out Einstein's field equation with cosmological
constant.
3) Gravitational gauge field theory is a renormalizable quantum
theory.
4) It gives out the same theoretical predictions on three
main classical tests as those of general relativity.
5) It can also predict the theoretical value of cosmological
constant.
\\

In one point of view, gravity is treated as physical interactions
in flat space-time. In this case, the physical space-time is
always flat. In another point of view, gravity is put into
the structure of space-time and space-time is curved
if the gravitational gauge field does not vanish. Two points
of view are based on different space-time, i.e., the first point
of view is based on physical space-time, and the second point
of view is based on gravitational gauge group space-time.
Two-points of view should give out the same physics, because there
is only one physics for gravity. In other
wards, they are equivalent. In the first point of view, gravity
is a kind of physical interactions, and in the second point of
view, gravity is the geometry of space-time. Because they are
equivalent, the nature of gravity is physics-geometry duality,
i.e., gravity is a kind of physical interactions
which has the characteristics of
geometry, it is also a geometry of space-time which is essentially
a kind of physical interactions.
\\

In general relativity, geometrical nature of gravity is the result
of equivalence principle, but equivalence principle is not the
necessary condition of the geometrical nature of gravity. From
mathematical point of view, the geometrical nature of gravity in
gauge theory of gravity originates from of geometrical nature of
gravitational gauge transformation, i.e., the geometrical nature
of translation transformation.
\\

{\bf \lbrack Acknowledgement \rbrack} The author wish to thank
Prof. F.W.Hehl, Prof. Germano Resconi,
Prof. Zhan Xu, Prof. Dahua Zhang and
Prof. Zhipeng Zheng for fruitful discussions on some
problems of gravity. \\


\begin{thebibliography}{99}

\bibitem{01} Isaac Newton, {\it Mathematical Principles of Natural Philosophy},
    (Camgridge University Press, 1934) .
\bibitem{02} Albert Einstein, Annalen der Phys., {\bf 49} (1916) 769 .
\bibitem{03} Albert Einstein, Zeits. Math. und Phys. {\bf 62} (1913) 225.
\bibitem{1} C.N.Yang, R.L.Mills, Phys Rev {\bf 96} (1954) 191 .
\bibitem{2} S.Glashow, Nucl Phys {\bf 22}(1961) 579 .
\bibitem{3} S.Weinberg, Phys Rev Lett {\bf 19} (1967) 1264 .
\bibitem{4} A.Salam, in Elementary Particle Theory, eds.N.Svartho .
Forlag, Stockholm,1968).
\bibitem{5} Albert Einstein, "Naeherungsweise Integration der
    Feldgleichungen der Gracitation", Preussische Akademie der
    Wissenschaften (Berlin) Sitzungsberichte, pg 688 (1916) .
\bibitem{b1} R.Utiyama, Phys.Rev.{\bf 101} (1956) 1597.
\bibitem{b2} A.Brodsky, D.Ivanenko and G. Sokolik, JETPH
    41 (1961)1307; Acta Phys.Hung. {\bf 14} (1962) 21.
\bibitem{b3} T.W.Kibble, J.Math.Phys. {\bf 2} (1961) 212.
\bibitem{b31} F. Mansouri and Lay Nam Chang,
        Phys. Rev. {\bf D13} (1976) 3192.
\bibitem{b32} S. W. MacDowell and F. Mansouri,
         Phys. Rev. Lett. {\bf 38} (1977) 739.
\bibitem{b33}F. Mansouri, Phys. Rev. {\bf D16} (1977) 2456.
\bibitem{b4} D.Ivanenko and G.Sardanashvily, Phys.Rep. {\bf 94}
    (1983) 1.
\bibitem{b5} F.W.Hehl, J.D.McCrea, E.W.Mielke and Y.Ne'eman
    Phys.Rep. {\bf 258} (1995) 1-171
\bibitem{b6} F.W.Hehl, P. Von Der Heyde, G.D.Kerlick, J.M.Nester
    Rev.Mod.Phys. {\bf 48} (1976) 393-416
\bibitem{wu1} Ning Wu, {\it Gauge Theory of Greavity}, hep-th/0109145
\bibitem{wu2} Ning Wu, (in preparation).
\bibitem{wu21} Ning Wu, Zhan Xu, Da-Hua Zhang, 
	{\it Differential Geometrical Formulation of Quantum
	Gauge Theory of Gravity } hep-th/0201131.
\bibitem{wu3} Ning Wu, Germano Resconi, Zhan Xu,
	Zhi-Peng Zheng, Da-Hua Zhang, Tu-Nan Ruan,
        {\it   Determination of Cosmological Constant from
                Gauge Theory of Gravity} gr-qc/0112063.
\bibitem{pdg} Particle Data Group, 2000,  Euro. Phys. J.
        {\rm C15}  1.
\bibitem{wu4} Ning Wu, {\it Schwarzschild-like Solution in
        Gauge Theory of Gravity} (in preparation).
\bibitem{wei} S.Weinberg, {\it Gravitation and Cosmology},
    (John Wiley, New York, 1972)




\end{thebibliography}
\end{document}